\documentclass[12pt]{article}
\title{Local anisotropy of space in a frame of reference co-moving with 
the Earth}
\author{Ll. Bel and A. Molina\thanks{Dep. de F\'\i sica Fonamental,
Universitat de Barcelona, Diagonal 647, Barcelona 08028 i Societat 
Catalana de F\'\i sica.}\\Lab. Gravitation et Cosmologie Relativistes. ESA 7065.\\
{\small Tour 22-12, 4, place Jussieu, 75252 Paris}
}
\date{}
\begin{document}
\maketitle

\begin{abstract}
We consider, in the framework of General Relativity, the linear approximation
of the gravitational field of the Earth taking into account its mass, its 
quadrupole moment, its shape and its diurnal rotation. 

We conclude that in the frame of reference co-moving with the Earth 
the local anisotropy of the space is of
the order of $10^{-12}-10^{-13}$ and could be observed.
\end{abstract}
\section{Introduction}

Three different types of experiments have been performed to test the
isotropy of space. Although this concept has always a geometrical
connotation what is really meant depends crucially on the type of
experiments being considered.

The first type of experiments are those of the type pioneered by
Michelson and Morley \cite{Michelson} and developed by Miller
\cite{Miller}, Joos \cite{Joos} and many others  using optical
interferometers. Jaseja and al. \cite{Jaseja} introduced a new setup
using a stabilized laser as a standard of frequency and a Fabry-Perot
as a standard of length, a technique that was used again by Brillet 
and Hall \cite{Brillet} who claim to have attained the best
sensitivity.

Historically the purpose of these experiments was to measure the
absolute velocity of the Earth, i.e. the velocity with respect to
that particular frame of reference for which the light propagation,
according to Newtonian Physics, was isotropic. In these experiments
$\alpha(t)=\triangle c/c$, where $\triangle c$ is the difference of
the round-trip speed of light along two perpendicular directions, say
south and east, on a horizontal plane, is measured as a function of time
$t$.  These experiments\,\footnote{From a careful reading of
\cite{Brillet} it follows that the upper limit is $10^{-13}$ if one
assumes that the anisotropy could have a local origin and 
$10^{-15}$ if one assumes that the anisotropy is necessarily a cosmic
one.} have consistently yield results for $\max. \alpha(t)$ 
ranging from $10^{-9}$ to $10^{-13}$. 

The second type of experiments was pioneered by Kennedy and Thorndike \cite{Kennedy} and is a variant of the
Michelson-Morley one in the sense that light travels along a single direction fixed with respect to the Earth and one attempts to
measure $\beta(t)=(c(t)-c(t_0))/c(t_0)$, where $c(t)$ is the round-trip speed of light along the fixed direction.      

The greatest value for $\alpha(t)$ obtained from the Michelson-Morley
experiment is already $10$ times smaller than the expected maximum
Newtonian value $10^{-8}$ and therefore these experiments are
considered to be good tests of Special Relativity.  In fact, as
Jaseja and al. have pointed out is not so obvious to decide what
these experiments test. More precisely  we ask here whether they
should be analyzed in the framework of Special Relativity, ignoring
or not the centrifugal force due to the Earth's rotation?, or should
this analysis have to take into account the Coriolis field and the
Newtonian gravitational field in the framework of General Relativity?

In the first case the Michelson-Morley experiment can be considered
to be a test of isotropy of space in a cosmic global sense. In the
second case the same experiment has to be considered as a test of the
theory of light propagation in the presence of local inertial and
gravitational fields. On the contrary the Kennedy-Thorndike
experiment can only be considered as a test of the isotropy of space
in the cosmic global sense. 

The distinction between the two points of view would be of
paramount importance if a future experiment of one of these types yields a
significant non-null result at some precision. The first point of
view would mean the failure of Special Relativity while the second
one does not mean that. It means that Special Relativity as a  global
model of space-time is not
applicable here and that gravitational fields interfere with light
already at a local level. 

The third type of experiments was pioneered by Drever and Hughes
\cite{Drever}. They test the variation of an atomic or nuclear level splitting due to a
fixed magnetic field as the orientation of this
field with respect to the stars changes due to the Earth's rotation and motion around the Sun.  These
experiments yield a limit to this type of anisotropy of the order of $10^{-
20}$, a limit that Lamoreaux and al.  \cite{Lamoreaux} pushed to
$10^{-22}$. Because the setup of these type of experiments is fixed with
respect to the Earth they can be compared with the Kennedy-Thorndike 
type, although the physics which is involved is quite different. 
But they cannot be compared to the Michelson-Morley type because this
one can discriminate between three cases: i) $\alpha(t)=0$, ii)
$\alpha=\,$constant$\,\not=0$ and iii) $\alpha(t)$ a function of time. 

In this paper we analyze the local anisotropy of space in the
neighborhood of the Earth's surface. The first section is based on the theory of Principal
transformations. The analysis of the anisotropy of space is based
on the exterior field only. The value of $\alpha$ predicted depends
on the latitude of the site where the experiment is performed and
three parameters close to the mean radius of the Earth.

The analysis of the second section is global but model dependent. We
have considered a model of the Earth with the observed oblateness but
assumed its density to be constant. This analysis depends also on
some heuristics that can be justified by the theory of Principal
transformations.

The third section contains a remainder of our interpretation of the
anisotropy of space and a numerical comparison of the results derived
from the approaches of the two preceding sections.

\section{The gravitational field of the Earth: A local approach.}

As an approximate model of the gravitational field of the Earth in the vacuum 
neighborhood of its surface we consider the following
line-element\,\footnote{We shall be using units such that 
$c=1$ and $G=1.$}:

\begin{equation}
\label {1.1}
ds^2=-(1-2U_G)dt^2+(1+2U_G)(dr^2+r^2(d\theta^2
+\sin^2\theta d\varphi^2))
\end{equation}
where $U_G$ is:

\begin{equation}
\label {1.2}
U_G=\frac{M}{r}\left(1+\frac{1}{2}\frac{J_2R^2(1-3\cos^2 \theta )}{r^2}\right)
\end{equation}
$M$ being the mass of the Earth, $J_2$ being the coefficient of its gravitational quadrupole
moment and $R$ the equatorial radius of the ellipsoidal surface. 
The ``polar" coordinates $(r, \theta, \varphi)$ have been chosen for
convenience and are related to a system of harmonic ``Cartesian" ones
$(x^i)$ by the usual formulas:

\begin{equation}
\label {1.4}
x^1=r\sin \theta \cos \varphi  , \quad x^2=r\sin \theta \sin \varphi  , \quad
x^3=r\cos \theta 
\end{equation}
 
Considering only those terms which are linear with respect to $M$
and/or $J_2$ the line-element \ref{1.1} is a solution of Einstein's
vacuum equations which actually reduce to the Newtonian equation:

\begin{equation}
\label {1.3}
\triangle U_G=0
\end{equation} 

To take into account the Earth's rotation $\Omega$ we select a new frame of
reference related to the preceding one by the transformation:

\begin{equation}
\label {1.5}
\varphi \rightarrow \varphi + \Omega t
\end{equation}

The line-element \ref{1.1} becomes:

\begin{equation}
\label {1.6}
ds^2=-(1-2(U_G+U_\Omega))dt^2
+2 A_\varphi dt d\varphi
+(1+2U_G)(dr^2+r^2( d\theta^2
+\sin^2 \theta d\varphi^2))
\end{equation}
where:

\begin{equation}
\label {1.9}
U_\Omega=\frac12\Omega^2 r^2\sin^2 \theta , \quad 
A_\varphi=\Omega r^2\sin^2 \theta 
\end{equation}

Introducing new ``cartesian coordinates" $(x^i)$ connected with the
new ``polar" coordinates by the same formula \ref{1.5} the
line-element \ref{1.6} becomes:

\begin{equation}
\label {1.10}
ds^2=-(1-2U)dt^2+2A_i dx^i dt+(1+2U_G)\delta_{ij}dx^i dx^j
\end{equation}
where:

\begin{equation}
\label {1.11}
U=U_G+U_\Omega, \\ 
\end{equation}
and the non zero components of $A_i$ are:

\begin{equation}
\label {1.12}
A_1=-\Omega y, \quad  A_2=\Omega x.
\end{equation}
From now on we shall neglect any terms proportional to powers of $\Omega$ greater than 2.

Let $\tau$ be the proper time along a time-like geodesic then, neglecting cubic and higher order terms in the velocities, 
the free fall of a test particle is described by the following
equations:

\begin{equation}
\label {1.7}
\frac{d^2x^k}{d\tau^2}+\hat\Gamma^k_{ij}\frac{dx^i}{d\tau}\frac{dx^j}{d\tau}  =\left(1+\delta_{ij}\frac{dx^i}{d\tau}\frac{dx^j}{d\tau}\right)\Lambda^k
+2\Omega^k{}_j\frac{dx^j}{d\tau} 
\end{equation}
where the Newtonian and Coriolis field are respectively:

\begin{equation}
\label {1.8}
\Lambda^i=\delta^{ij}\partial_j(U_G+U_\Omega), \quad
\Omega^i{}_j=\frac12\delta^{is}(\partial_s A_j-\partial_j A_s)
\end{equation}
and:

\begin{equation}
\label {1.13}
\hat\Gamma^k_{ij}=\delta^k_j\partial_i U_G+\delta^k_i\partial_j U_G
-\delta_{ij}\delta^{ks}\partial_s U_G
-(\Omega_i{}^k\Omega_{jl}+\Omega_j{}^{k}\Omega_{il})x^l
\end{equation}	
Notice that, whether $\Omega$ is taken into account or
not, Eqs. \ref{1.7} are meaningless without clarifying the meaning of the
coordinates being used. Notice also that the newly defined ``cartesian"   
coordinates $x^i$ are no longer harmonic coordinates.

The symbols $\hat\Gamma^k_{ij}$ in \ref{1.7} are the Christoffel symbols
corresponding to the Riemannian metric:

\begin{equation}
\label {1.15}
d\hat s^2=((1+2U_G)\delta_{ij}+
\Omega_{ik}\Omega_{jl}x^k x^l)dx^i dx^j
\end{equation}
which is the quotient of the space-time metric \ref{1.6} by the
Killing congruence of the frame of reference.

This object is usually interpreted as defining the ``geometry" of
the space for the frame of reference being considered. This
interpretation is correct or not depending on the meaning that one
attaches to this word. It is certainly correct in a technical,
current and accepted sense meaning a well established branch of
mathematics.  It is not correct if by ``geometry of space'' we mean
that particular Riemannian geometry which allows the
description of rigid objects which can be displaced, or compared when
they occupy different positions in space. This point has been
strongly emphasized by Poincar\'e with profound insight
\cite{Poincare}, and requires the geometry of space to have constant
curvature \cite{Cartan}, the simplest case being of course the case
in which this geometry is flat. 

	If the object \ref{1.15} is not the metric of space, what is
it?. Before reminding a new interpretation which was first proposed in
\cite{Bel90} we are going to state and solve the problem of constructing the
Principal transformation of the Riemannian metric \ref{1.15}, this being a
concept introduced by one of us (Ll.B) in \cite{Bel96}. 

Using polar coordinates we have:

\begin{equation}
\label {1.17}
d\hat s^2=(1+2U_G)(dr^2+r^2 d\theta^2
+r^2\sin^2 \theta (1+2U_\Omega)d\varphi^2))
\end{equation} 

The principal directions, i.e. the eigen-vectors of $\hat R_{ij}$,
the Ricci tensor of this metric with respect to $\hat g_{ij}$ are to
the required approximation proportional to:

\begin{equation}
\label {1.19}
n_{1i}=(1, J_2 f, 0), \quad  n_{2i}=(-J_2 f, 1, 0), \quad 
n_{1i}=(0, 0, 1)
\end{equation}
where we have used the euclidean orthonormal co-basis $(dr, rd\theta,
r\sin \theta d\varphi)$ and where:

\begin{equation}
\label {1.22}
f=-4\frac{R^2}{r^2}\left(1-q\frac{r^5}{R^5}\right)\sin\theta\cos\theta  \quad \hbox{with} \quad 
q=\frac14\frac{\Omega^2 R^3}{M J_2}
\end{equation}
The vector $n_1$ is up to the required
approximation, as a short calculation shows, collinear with the 
gradient of $U$ pointing outwards. The vector $n_2$ is therefore that vector lying in the
meridian of the point being considered which is orthogonal to the
vertical pointing south, and $n_3$ is the the vector lying in the
corresponding parallel and pointing east. 

Using the preceding vectors as a basis of an orthogonal
decomposition the line-element \ref{1.17} can be written:

\begin{equation}
\label {1.20}
d\hat s^2=(\hat\theta^1)^2+(\hat\theta^2)^2+(\hat\theta^3)^2
\end{equation}
where:

\begin{equation}
\label {1.21}
\hat\theta^1=(1+U_G)(dr+J_2 f rd\theta), \,\,
\hat\theta^2=(1+U_G)(-J_2 f dr+ rd\theta), \,\,
\hat\theta^3=(1+U)r\sin \theta d\phi
\end{equation}
By definition, in this particular case, a Principal transform of
\ref{1.20} is a metric which can be written as:

\begin{equation}
\label {1.23}
d\bar s^2=c_1^2(\hat\theta^1)^2+c_2^2(\hat\theta^2)^2+c_3^2(\hat\theta^3)^2
\end{equation}
the three functions $c_i(x^j)$ being chosen such that:

i) the metric \ref{1.23} is euclidean, i.e. such that:

\begin{equation}
\label {1.24}
\bar R^i_{jkl}=0
\end{equation}

ii) the following ``quo-harmonicity condition" is satisfied:

\begin{equation}
\label {1.25}
(\bar \Gamma^i_{jk}-\hat \Gamma^i_{jk})\hat g^{jk}=0 
\end{equation}
where $\bar\Gamma^i_{jk}$ are the
Christoffel symbols of the line-element \ref{1.23}, and 
$\hat g^{jk}$ is the inverse matrix of $\hat g_{jk}$.

Notice that both groups of conditions are tensor equations and
therefore the solutions $c_i(x^j)$ are scalar
functions independent of the system of coordinates being used, as
long as this system is adapted to the Killing congruence defining the
static frame of reference.

Integrating \ref{1.24} and \ref{1.25} we have obtained to the required
approximation:

\begin{eqnarray*}
\label {1.26}
c_1=1-\frac{M}{r}\left(1+\frac15\frac{R_1^2}{r^2}\right)+
\frac{M J_2 R^2}{r^3}\left(6+\frac{6}{5}\frac{R^2}{r^2}-\left(5+\frac{9}{5}\frac{R^2}{r^2}
\right)\sin^2\theta\right)\\
+\Omega^2 r^2\left(-\frac{27}{5}+\frac{3}{10}\frac{R^2}{r^2}
+\left(\frac{15}{2}-\frac{3}{10}\frac{R^2}{r^2}\right)\sin^2 \theta\right)
\end{eqnarray*}

\begin{eqnarray*}
\label {1.27}
c_2=1-\frac{M}{2r}\left(3-\frac{1}{5}\frac{R_2^2}{r^2}\right)+
\frac{M J_2 R^2}{r^3}\left(1-\frac{3}{5}\frac{R^2}{r^2}-\left(\frac{19}{4}-\frac{21}{20}\frac{R^2}{r^2}\right)\sin^2\theta\right) \\
+\Omega^2 r^2\left(-\frac{19}{5}
+\left(\frac{13}{2}+\frac{3}{10}\frac{R^2}{r^2}\right)\sin^2 \theta\right)
\end{eqnarray*}

\begin{eqnarray*}
\label {1.28}
c_3=1-\frac{M}{2r}\left(3-\frac{1}{5}\frac{R_2^2}{r^2}\right)+
\frac{M J_2 R^2}{r^3}\left(1-\frac{3}{5}\frac{R^2}{r^2}-\left(\frac{9}{4}+\frac{3}{4}\frac{R^2}{r^2}\right)\sin^2\theta\right) \\
+\Omega^2 r^2\left(-\frac{19}{5}+4 \sin^2 \theta\right)
\end{eqnarray*}
where $R_1$ and $R_2$ are in this approach two model-dependent free parameters.
They should be determined by matching the preceding results with the 
corresponding ones in the interior of the Earth. This would require a
global ${\cal C}^2$ model including the interior gravitational field
and therefore also a ${\cal C}^0$ behavior of the
density of the Earth. To deal with a discontinuous density requires
to adopt a different approach. This is done in the next section.

\section{The gravitational field of the Earth: A global approach.}
To begin with we consider in this section a model of the Earth with constant density taking into account its oblateness.  More precisely we shall assume
that the surface of the Earth $\Sigma$ is $r^2=a^2(1+\lambda^2\sin^2 \theta)$, where $a$ is the polar radius, and $\lambda=\sqrt{R^2-a^2}/R$ the eccentricity, $R$ as before is the equatorial radius.

An elementary integration of the corresponding Poisson equation yields up
to order $J_2$ the
Newtonian potential:
\begin{equation}
\label{2.1}
U_G=\left\{\begin{array}{lr}
{\displaystyle \frac{M}{r}\left(1+\frac{J_2}{2}\frac{R^2}{r^2}(1-3\cos^2\theta)\right)} & \mbox{outside}\\[1em]
{\displaystyle \frac{M}{R}\left(\frac32-\frac12\frac{r^2}{R^2}+\frac{J_2}{2}\left(\frac52-\frac32\frac{r^2}{R^2}-3\frac{r^2}{R^2}\cos^2\theta\right)\right)}&\mbox{inside}
\end{array}
\right.
\end{equation}
which is of class ${\cal C}^1$ across $\Sigma$, where $J_2=\lambda^2/5$.

Taking into account also the diurnal rotation of the Earth, the gravitational field in a comoving frame of reference is then described by the line element \ref{1.10}, where
now $U_G$ is given by \ref{2.1} instead of \ref{1.2} and correspondingly we have to make the same substitution in \ref{1.17}.

On the surface of the Earth $\Sigma$ the Ricci tensor of the metric 
\ref{1.17} is discontinuous as it is the density and therefore the 
principal directions derived from the exterior field do not coincide 
with those derived from the interior one. This prevents the existence 
for this metric of a global principal transformation continuous across 
$\Sigma$.

When a  principal transformation exists then from \ref{1.25} it follows 
that cartesian coordinates of \ref{1.23}, i. e. those with 
$\bar\Gamma^i_{jk}=0$ are quo-harmonic coordinates of \ref{1.17}, i. e. 
they are such that $\hat\Gamma^i_{jk}\hat{g}^{ij}=0$. The approach will 
then be heuristically
justified by this result. We shall obtain a global system of 
quo-harmonic coordinates, unique up to euclidean transformations, 
and we shall consider the anisotropy  of the space to be the 
anisotropy of \ref{1.17} with respect to the euclidean metric 
which in this system of coordinates has components
$\bar{g}_{ij}=\delta_{ij}$.\footnote{This point of view was already
used in \cite{ABMM} where harmonic coordinates of the space-time were 
considered instead of the quo-harmonic coordinates of the space. It 
could be considered also in connection with rotating Fermi coordinates 
based on the world-line of the center of the Earth. These two coordinate 
dependent models are heuristic as the one presented in this section. 
The stronger legitimacy of the latter comes from being ackin with the 
method of the first section.}

Under an infinitesimal coordinate transformation
$$x^i=z^i+\zeta^i(z^j)$$
where $\zeta_i$ are of the order appropriate to the approximation that we consider, the Riemannian metric \ref{1.15} becomes:
$$\hat{g}^\prime_{ij}=\hat{g}_{ij}(x^k(z^l))+
\frac{\partial\zeta_j}{\partial z^{i}}+\frac{\partial\zeta_i}{\partial z^{j}}.$$

The new coordinates $z_i$ will be quo-harmonic if: \cite{Salas}
\begin{equation}
\label{eq2.2}
\bigtriangleup
 \zeta_k(z^i)=\frac{\partial U}{\partial z^k}.
\end{equation}
We can solve this equation splitting the functions $\zeta_k$ in their
gravitational and rotational part $\zeta_k=\zeta_k^G+\zeta_k^\Omega$.
Then 
\begin{equation}\label{eq2.3}
\bigtriangleup\zeta_k^G=\frac{\partial U_G}{\partial z^k},
\end{equation}
\begin{equation}\label{eq2.4}
\bigtriangleup\zeta_k^\Omega=-2\Omega_{kj}\Omega^j{}_l z^l.
\end{equation}

The solution for $\zeta_k^G$ can found by integrating the Poisson equation \ref{eq2.3} because the ``density" 
$\partial U_G/\partial z^k$ is known. Requiring the gradient of $\zeta_k^G$ to be zero at infinity. In the new polar coordinates, i.e. coordinates $(r, \theta, \varphi)$ such that:
\begin{equation}
\label {1.4}
z^1=r\sin \theta \cos \varphi  , \quad z^2=r\sin \theta \sin \varphi  , \quad
z^3=r\cos \theta 
\end{equation}
we obtain:
\begin{equation}
\label{eq2.5}
\zeta_r^G=\left\{\begin{array}{lr}
M\left( 
{\displaystyle 
\frac12 -\frac{1}{10}\frac{R^2}{r^2} + J_2\frac{R^2}{r^2}\left(\frac{3}{14} 
\frac{R^2}{r^2}+\frac14\left(1 -\frac97\frac{R^2}{r^2}\right)\sin^2\theta\right)}\right) & \mbox{outside}\\[1.5em]
 M {\displaystyle\frac{r}{R} 
\left(
\frac12 -\frac{1}{10}\frac{r^2}{R^2}+J_2 
\left( \frac34-\frac{15}{28}\frac{r^2}{R^2}-\frac12\left(1-\frac67\frac{r^2}{R^2}\right)\sin^2\theta\right)\right)}&
\mbox{inside}
\end{array}\right.
\end{equation}

\begin{equation}
\label{eq2.6}
\zeta_\theta^G=\left\{\begin{array}{lr}
- {\displaystyle \frac{M}{2}J_2\frac{R^2}{r}\sin\theta
\cos\theta\left(1 -\frac37\frac{R^2}{r^2} \right)}
 &\mbox{outside} \\[1.5em]
- {\displaystyle \frac{M}{2}J_2\frac{r^2}{R}\sin\theta
\cos\theta\left(1 -\frac37\frac{r^2}{R^2} \right)} &\mbox{inside}
\end{array}\right.
\end{equation}

The rotational part can be obtained very easily from an inspection of
the explicit expression of  \ref{eq2.4}. Demanding the appropriate symmetry and the correct limit on the axis of symmetry we obtain:

\begin{equation}
\label{eq2.7}
\zeta_r^\Omega={\displaystyle\frac14 \Omega^2r^3\sin^4\theta}\quad
\zeta_\theta^\Omega={\displaystyle\frac14\Omega^2 r^4\sin^3\theta\cos\theta} 
\end{equation}
Writing the metric \ref{1.15} as $\hat{g}_{ij}=\delta_{ij} +\hat{h}_{ij}$ and using polar coordinates we obtain for $\hat{h}_{ij}$:
\begin{equation}
\label{eq2.8}\begin{array}{l}
\hat{h}_{rr}={\displaystyle 
2\frac{M}{r}\left(1+\frac15\frac{R^2}{r^2}+J_2\frac{R^2}{r^2}\left(-1-\frac67\frac{R^2}{r^2}+\left(1+\frac97\frac{R^2}{r^2}\right)\sin^2\theta
\right)\right)+} \\[1em]\hfill{\displaystyle\frac32\Omega^2r^2
\sin^4\theta},\\[1em]
\hat{h}_{\theta\theta}={\displaystyle Mr\left(3- 
\frac15\frac{R^2}{r^2}-J_2\frac{R^2}{r^2} 
\left(-3+\frac67\frac{R^2}{r^2}+ \left(\frac{11}{2}-
\frac32\frac{R^2}{r^2}\right)\sin^2\theta\right)\right)}+
\\[1em]\hfill {\displaystyle\frac32\Omega^2r^4\sin^2\theta
\cos^2\theta},\\[1em]
\hat{h}_{\phi\phi}={\displaystyle Mr\sin^2\theta\left(3- \frac15
\frac{R^2}{r^2}+J_2\frac{R^2}{r^2}\left(-3+\frac67\frac{R^2}{r^2}+
\left(\frac92-\frac{15}{14}\frac{R^2}{r^2}\right)\sin^2\theta\right)
\right)}+\\[1em]\hfill{\displaystyle\frac32\Omega^2r^4\sin^4\theta}, \\[1em]
\hat{h}_{r\theta}={\displaystyle
\sin\theta\cos\theta\left(2MJ_2\frac{R^2}{r^2}
\left(1-\frac67\frac{R^2}{r^2}\right)+
\frac32\Omega^2r^3\sin^2\theta\right)}.
\end{array}
\end{equation}

Making more precise what we said before we define the eigen-values and the principal directions of the anisotropy of the metric \ref{1.17} as the scalars $c_a$ and the vectors $n^i_a$ solutions of the algebraic 
equations: $$(\hat{g}_{ij}-c_a\delta_{ij})n^j_a=0.$$
with $\hat{h}_{ij}$ given above by \ref{eq2.8}. To the required approximation the scalars $c_a$ are:  
\begin{equation}
\label{eq2.9}
c_1=1+\hat{h}_{rr}, \quad c_2=1+\frac{\hat{h}_{\theta\theta}}{r^2}, \quad c_3=1+\frac{\hat{h}_{\phi\phi}}{r^2\sin^2\theta}
\end{equation}
and the components of the principal directions in the euclidean orthonormal co-basis $(dr, rd\theta, r\sin \theta d\varphi)$ are proportional to:
\begin{equation}
\label{eq2.10}
n_{1i}=(1,f(r,\theta),0),\quad  n_{2i}=(-f(r,\theta),1,0),\quad 
n_{3i}=(0,0,1)
\end{equation}
where:
$$f(r,\theta)=\frac{-\hat{h}_{r\theta}}{M(1-3R^2/(5r^2))}.$$

\section*{Conclusion}

In section 2 we exhibited the anisotropy of the riemannian metric \ref{1.17} with
respect to an intrinsically associated euclidean metric. This
association depends on three model dependent parameters $R_1, R_2$ and
$R$. If the quadrupole moment were zero the
three parameters would coincide and assuming a model of the Earth
with constant density their value would be its radius. Therefore
we can safely guess that the three parameters will be close to each
other and of the order of the mean radius of the
Earth. It is worthwhile to notice that on any
horizontal plane, i.e orthogonal to the gradient of $U$ the
anisotropy depends only on $R$. And in particular if $r=R\approx R_1\approx R_2$ we get:

\begin{equation}
\label {1.29}
\alpha=c_2-c_3=-\frac15\left(\frac{11MJ_2}{R} -14\Omega^2R^2\right)\sin^2\theta 
\end{equation}   
Assuming the following numerical values: $M=0.00444$ $R=6378164$ 
and $\Omega=2.434\times 10^{-13}$, the three of them measured in
meters, and $J_2=0.0010826$ dimensionless, at a latitude of $45^\circ$
we obtain:

\begin{equation}
\label {1.30}
\alpha=c_2-c_3=2.5\times 10^{-12}
\end{equation} 

In section 3 we constructed a globally ${\cal C}^1$ model of the gravitational field
of the Earth assuming it to be of constant density
and having an ellipsoidal shape.\footnote{This model of the Earth would
yield a value of $\lambda^2=0.00541.$ while the observed value is $\lambda^2=0.00674$} Then we obtained a global system of
quo-harmonic coordinates of the space.  It follows
from that that we can associate with the global field of the Earth an 
euclidean  	
metric with components $\bar g_{ij}=\delta_{ij}$ in
this unique system of quo-harmonic coordinates. This is again a model dependent
but intrinsic association.  

With this model the anisotropy in the horizontal plane, for $r=R$ is:

\begin{equation}
\label {1.31}
\alpha=c_2-c_3=\left(-\frac{4}{7}\frac{MJ_2}{R}+\frac32\Omega^2 R^2\right)\sin^2\theta
\end{equation}
and with $J_2$, $M$, $R$, $\Omega$ and $\theta$ as 
before, this formula yields:

\begin{equation}
\label {1.32}
\alpha=c_2-c_3=6.2\times 10^{-13}
\end{equation} 

The two results in \ref{1.30} and \ref{1.32} differ, although by less of a factor of
$10$. The safe conclusion that we can draw from our precedent
analysis is that it predicts an anisotropy on the horizontal plane 
in the range $10^{-12}-10^{-13}$. \footnote{The harmonic coordinate method and the 
Fermi coordinate one yield values in the same range.}  It is however important to remind
that the theory of Principal transformations tell us that for a model
of the gravitational field of the Earth of class ${\cal C}^2$ the two
methods would yield the same result.

The important question is now the following: how could this
anisotropy show up?. Because the anisotropy is not due to any cosmic
effect but is due to the gravitational field of the Earth and to its
diurnal rotation and because it is relative to a co-moving frame of
reference, no experiment of the Kennedy-Thorndike or 
Hughes-Drever type could reveal it.
If instead, as it was first proposed in \cite{Bel90}, \cite{ABMM} and \cite{BMM}, we renounce
to interpret the metric \ref{1.17} with components $\hat g_{ij}$ as
the geometry of space, a role which belongs to the euclidean metric
$\bar g_{ij}$, and we interpret the eigen-values of the first 
of this objects with respect to the second as the velocities of light along the
principal directions of an anisotropic propagation, then the experiments of the Michelson-Morley type should reveal such anisotropy. 

As a matter of fact Brillet and Hall, the authors of the most recent
repetition of this experiment obtained a raw result of the order of
$2.1\times 10^{-13}$ at the latitude of $+40^\circ$. If confirmed our
model could provide an explanation for this result. In our opinion
is an urgent task that somebody repeats this experiment using their 
setup or any other more appropriate variant.
\section*{Acknowledgments}
We want to acknowledge many useful discussions with C. L\"ammerzahl and 
P. Teyssandier and we thank them for a very careful reading of the manuscript.

One of us A. M.  wishes to acknowledge the hospitality of the Laboratoire
de Gravitation et Cosmologie Relativistes (UPRESA 7065) and the 
contracts No.~PB96-0384 from DGES, ``Ministerio de Educaci\'on y 
Cultura" and No.~1996SGR-00048 from CURG, ``Generalitat de Catalunya" 
and specially a grant from the ``Programa sectorial de formaci\'on de 
profesorado" of the ``Ministerio de educaci\'on y cultura".


\begin{thebibliography}{99}
\bibitem{Michelson} A. A. Michelson and E. W. Morley, Am. J. Sci. {\bf 34}, 333 (1887).
\bibitem{Miller} D. C. Miller, Rev. Mod. Phys. {\bf 5}, 203 (1933).
\bibitem{Joos} M G. Joss, Ann. Phys. {\bf 7}, 385 (1930). 
\bibitem{Jaseja} T. S. Jaseja, A. Javan, J. Murray and C. H. Townes, Phys. Rev. A {\bf 133}, 1221 (1964).
\bibitem{Brillet} A. Brillet and J.L. Hall,  Phys. Rev. Let. {\bf
42}, 549  (1979).
\bibitem{Kennedy} R. J. Kennedy and E. Thorndike, Phys. Rev. {\bf
42}, 400 (1932).
\bibitem{Drever} V. W. Hughes, H. G. Robinson and V. Beltran-Lopez, Phys.
Rev. Lett. {\bf 4}, 342, (1960). R. W. P. Drever, Phylos. Mag. {\bf 6}, 683
(1961).
\bibitem{Lamoreaux} S. K. Lamoreaux, J. P. Jacobs, B. R. Heckel, F. J. Raab
and E. N. Fortson, Phys. Rev. Let. {\bf 25}, 3125 (1986).
\bibitem{Poincare} H. Poincar\'e ``La science et l'hypoth\`ese" Chapitre IV.
\bibitem{Cartan} E. Cartan, Le\c{c}ons sur la g\'{e}om\'{e}trie des espaces 
de Riemann, Gautiers--Villards (1951).
\bibitem{Bel90} Ll. Bel, Recent developments in gravitation ed. 
E. Verdaguer,  J. Garriga, J. Cespedes, (World Scientific Pub. 
Co., 1990).
\bibitem{Bel96} Ll. Bel, Gen. Rel. and Grav. {\bf 28}, 1139 (1996).  
\bibitem{Salas}Ll. Bel,  Relativity in General, J. D\'\i az and M. Lorente editors, Editions Fronti\`eres, p. 47 (1994).
\bibitem{ABMM} J. M. Aguirregabiria, Ll. Bel, J.
Mart\'\i n and A. Molina, Proceedings of the 1991 Relativity Meeting, Bilbo ``Recent Developments in Gravitation", p. 61-79 
 Ed. A. Feinstein and J. Ib\'a\~{n}ez, World Scientific (1992).
\bibitem{BMM}Ll. Bel, J. Mart\'\i n and A. Molina, Jour. Phys. Soc. Japan, 
{\bf 63}, 4350 (1994).
\end{thebibliography}
\end{document}